\documentclass[12pt]{article}
\makeatletter
\newcommand{\singlespacing}{\let\CS=\@currsize\renewcommand{\baselinestretch}
{1.0}\tiny\CS}
\newcommand{\doublespacing}{\let\CS=\@currsize\renewcommand{\baselinestretch}
{1.5}\tiny\CS}

\def\sech{\mathop{\rm sech}\nolimits}
\def\cosech{\mathop{\rm cosech}\nolimits}

\begin{document}

\title{
Effective-Mass Schr\"{o}dinger Equation and Generation of Solvable Potentials}
\author{B. Bagchi$^{a,}$\thanks{\bf E-mail:bbagchi123@rediffmail.com}\ , \ P.
Gorain$^{a,}$\thanks{\bf E-mail:psgorain@rediffmail.com}\ , \ C.
Quesne$^{b,}$\thanks{\bf E-mail:cquesne@ulb.ac.be} \ and R.
Roychoudhury$^{c,}$\thanks{\bf E-mail:raj@isical.ac.in} \\
\\
$^a$ Department of Applied Mathematics, University of Calcutta, \\ 92 Acharya Prafulla
Chandra Road, Kolkata 700 009, India \\ \\
$^b$ Physique Nucl\'{e}aire Th\'{e}orique et Physique Math\'{e}matique, \\
Universit\'{e} Libre de Bruxelles, Campus de la Plaine CP229, \\ Boulevard du
Triomphe, B-1050 Brussels, Belgium \\ \\
$^c$ Physics and Applied Mathematics Unit,\\ Indian Statistical Institute, \\ Kolkata
700 108, India}

\date{}

\maketitle

 \vspace*{0.5cm}

{\centerline{\bf ABSTRACT}}

\vspace{0.2cm}

\thispagestyle{empty}

\setlength{\baselineskip}{18.5pt}

A one-dimensional Schr\"odinger equation with position-dependent effective mass in
the kinetic energy operator is studied in the framework of an $so(2,1)$ algebra. New
mass-deformed versions of Scarf II, Morse and generalized P\"oschl-Teller potentials
are obtained. Consistency with an intertwining condition is pointed out.

%
%
\newpage

Significant attention has been focussed on the issue of
position-dependent-effective-mass (PDEM) quantum Hamiltonians and their impact on the
construction of soluble quantum systems. Interest in PDEM stems from its physical
relevance in problems of compositionally graded crystals [1], quantum dots [2], liquid
crystals [3], etc., where the need for a varying mass has long been felt. Indeed the
appearance of PDEM is well known in the energy density functional approach to the
nuclear many-body problem in the context of nonlocal terms of the accompanying
potential [4, 5, 6]. Exact solutions of the PDEM Schr\"odinger equation have also
been reported by extending the methods of coordinate transformation and
supersymmetric quantum mechanics [7--20]. In particular, for the free-particle case we
have recently found [21] by exploiting the intertwining relation that for an
appropriate choice of the mass function the problem can be completely solved leading
to normalizable bound states.

The purpose here is to study the PDEM within an $so(2,1)$ algebra and obtain new
mass-deformed solutions of the underlying functions characterizing the $so(2,1)$,
which are natural counterparts of those in the constant-mass case. We also realize
that a class of solutions exists for which the effective potential induced by
$so(2,1)$ coincides with the one provided by a first-order intertwining condition.

The PDEM kinetic energy operator has a wide range of forms. In the following we adopt
von Roos' scheme [22], which has the advantage of a builtin Hermiticity. It is
given by
$$ \widehat T = \frac {1}{4} \left[ m^\alpha (x) \hat p m^\beta (x) \hat p m^\gamma
(x) + m^\gamma (x) \hat p m^\beta (x) \hat p m^\alpha (x) \right], \eqno (1) $$
where
$\hat p \left(\equiv -i\hbar \frac {d}{dx} \right)$ is the momentum operator, $m(x)$
is the position-dependent mass and the parameters $\alpha, \beta, \gamma$ are tied
by the condition $\alpha + \beta + \gamma = -1$. We have shown elsewhere [21] that
if we set $m(x) = m_0 M(x)$, $M(x)$ being dimensionless, and use the identity
$$\begin{array}{l} M^\alpha \frac {d}{dx} M^\beta \frac {d}{dx} M^\gamma + M^\gamma
\frac {d}{dx} M^\beta \frac {d}{dx} M^\alpha = 2\frac {d}{dx}
\frac {1}{M} \frac {d}{dx} - (\beta +1) \frac {M^{\prime\prime}}{M^2}\\ \\ + 2\left[
\alpha (\alpha +
\beta + 1) + \beta + 1\right] \frac {M^{\prime^2}}{M^3},\\
\end{array} \eqno (2)
$$ the primes denoting derivatives with respect to the variable $x
\in (-\infty, \infty)$, then the ambiguity parameters
$\alpha, \beta, \gamma$ get shifted to an overall effective potential energy term.
With $\hbar = 2m_0 = 1$, we get, for a given potential $V(x)$, a modified
Schr\"odinger equation that reads
$$ H\psi (x) \equiv \left[ -\frac {d}{dx} \frac {1}{M(x)} \frac {d}{dx} + V_{\rm eff} (x)
\right] \psi (x) = E \psi (x), \eqno (3) $$ in which $V_{\rm eff} (x)$ also depends upon
$M(x)$ and its derivatives:
$$ V_{\rm eff} (x) = V(x) + \frac {1}{2} (\beta + 1) \frac {M^{\prime\prime}}{M^2} -
\left[ \alpha (\alpha + \beta + 1) + \beta + 1\right] \frac {M^{\prime^2}}{M^3}. \eqno
(4) $$

Consider now the generators of the $so(2,1)$ algebra represented by
$$\begin{array}{lcl}
\displaystyle J_0 & = & \displaystyle -i \frac {\partial}{\partial \phi}, \\ \\
\displaystyle J_\pm & = & \displaystyle e^{\pm i\phi} \left[ \pm
\frac {1}{\sqrt {M}} \frac {\partial}{\partial x} + F(x) \left( i
\frac {\partial}{\partial \phi} \mp \frac {1}{2} \right) + G(x)
\right], \end{array} \eqno (5) $$ where $\phi$ is an auxiliary variable and we restrict
to $M(x) > 0$. The commutation relations of $so (2,1)$, namely $[J_+, J_-] = -2J_0$,
$[J_0, J_\pm] = \pm J_\pm$, imply that the functions $F$ and $G$ be constrained by
the equations
$$ F^\prime = \sqrt {M} (1-F^2), \qquad \ G^\prime = -\sqrt {M} F G. \eqno (6) $$

With
$$\begin{array}{lcl}
\displaystyle J_0 | k \mu > & = & \displaystyle \mu | k\mu >, \\[0.2cm]
\displaystyle J^2 | k \mu > & = & \displaystyle k (k-1) | k\mu >,
\end{array} \eqno (7) $$ where $|k \mu\rangle = \chi_{k\mu} (x) e^{i\mu
\phi}$ are basis functions appropriate to an $so (2,1)$ irreducible representation of
the type $D_k^+$ and $\mu$ takes on values $k$, $k+1$, $k+2, \ldots$, the Casimir
operator $J^2 = J_0^2 \mp J_0 - J_\pm J_\mp$, when expanded, gives
$$ \left[ - \frac {1}{\sqrt {M}} \frac {d}{dx} \frac {1}{\sqrt M}
\frac {d}{dx} + V_\mu \right] \chi = -\left( k - \frac {1}{2}
\right)^2 \chi, \eqno (8) $$ in which $V_\mu$ describes a one-parameter family of
potentials
$$ V_\mu = \frac{1}{\sqrt{M}} \left[\left( \frac {1}{4} - \mu^2 \right) F^\prime +
2\mu G^\prime\right] + G^2.
\eqno (9) $$ From (8), it follows that the functions $\chi(x) \equiv \chi_{k\mu}(x)$
are the eigenfunctions of different Hamiltonians but conform to the same energy level.

Equation (8) is in a direct one-to-one correspondence with the von-Roos-generated
PDEM form (3) if we transform $\chi (x)
\rightarrow [M(x)]^{-1/4} \psi (x)$. We have
$$ \left[ - \frac {d}{dx} \frac {1}{M(x)} \frac {d}{dx}
+ \frac {M^{\prime\prime}}{4M^2} - \frac {7M^{\prime^2}}{16M^3} + V_\mu \right] \psi
= - \left( k - \frac {1}{2} \right)^2 \psi, \eqno (10) $$ suggesting the
identifications
$$ V_{\rm eff} (x) = \frac {M^{\prime\prime}}{4M^2} - \frac {7M^{\prime^2}}{16M^3}
+ V_\mu, \eqno (11) $$
$$ E = - \left( k - \frac {1}{2} \right)^2, \eqno (12) $$ along with
$$ V(x) = \left[ \alpha (\alpha + \beta + 1) + \beta + \frac {9}{16}\right]
\frac {M^{\prime^2}}{M^3} - \frac {1}{4} (2\beta + 1)
\frac {M^{\prime\prime}}{M^2} + V_\mu \eqno (13) $$ on using the expression (4) for
$V_{\rm eff} (x)$.

The set of equations (10) -- (13) extends the realization of $so(2,1)$ to the PDEM
case. In particular, Eq.~(13) states that for a given potential $V(x)$ placed in a
suitable mass environment, $so (2,1)$ as a potential algebra can be realized for it
characterized by $V_\mu$ and supporting the same set of energy eigenvalues (12). Of
course, in the constant-mass case, $V(x)$ reduces to $V_\mu$, which is as it should be.

A couple of observations are in order:

\noindent (a) A change of variable
$$ u(x) = \int^x \sqrt {M(t)}\, dt \eqno (14) $$ allows one to avoid explicit presence
of $\frac {1}{\sqrt {M}}$ factor in the generators (5). Consequently the conditions
(6) read
$$ \dot {\displaystyle\mathop F^\sim} = 1- {\displaystyle\mathop F^\sim}^2, \qquad
\dot {\displaystyle\mathop G^\sim} = - {\displaystyle\mathop F^\sim}
{\displaystyle\mathop G^\sim}, \eqno (15) $$ where the overhead dot indicates a
derivative with respect to the variable
$u(x)$ and $({\displaystyle\mathop F^\sim}, {\displaystyle\mathop G^\sim})$ are
transformed $(F, G)$ under (14). The forms (15) are similar to those for the
conventional constant-mass case, for which it is known [23,24] that there exist at
most three classes of realizations of the functions ${\displaystyle\mathop F^\sim}$ and
${\displaystyle\mathop G^\sim}$ depending on the sign of
${\displaystyle\mathop \omega^\sim} = \Bigl({\displaystyle\mathop
F^\sim}^2-1\Bigr)\Big/{\displaystyle\mathop G^\sim}^2$:
$$ {\displaystyle\mathop \omega^\sim} =-\frac {1}{b^2} < 0 :
\qquad{\displaystyle\mathop F^\sim} (u) =
\tanh (u-c), \qquad {\displaystyle\mathop G^\sim} (u) = b \sech(u-c),$$
$$ {\displaystyle\mathop \omega^\sim} = 0 :\qquad {\displaystyle\mathop F^\sim} (u)
= \pm 1, \qquad {\displaystyle\mathop G^\sim} (u) = b e^{\mp u}, \eqno (16) $$
$$ {\displaystyle\mathop \omega^\sim} = \frac{1}{b^2} > 0 :\qquad
{\displaystyle\mathop F^\sim} (u) =
\coth (u-c), \qquad {\displaystyle\mathop G^\sim} (u) = b \cosech (u-c), $$ where
$b$ and
$c$ are constants. Hence there would be three classes of realizations for (5) too,
depending on the sign of $\omega = (F^2-1)/G^2$ according to which
$$ \omega = -\frac {1}{b^2} < 0 :\qquad F(x) = \tanh \left[ u(x) - c\right], \qquad G(x)
= b\sech [u(x) - c], $$
$$ \omega = 0 :\qquad F(x) = \pm 1, \qquad G(x) = b e^{\mp u(x)}, \eqno(17) $$
$$ \omega = \frac {1}{b^2} > 0 :\qquad F(x) = \coth \left[ u(x) - c \right],
\qquad G(x) = b \cosech \left[ u(x) - c \right],$$ where
$u$ is known as a function of $x$ from (14) for some mass function
$M(x)$. For instance, a plausible choice of the latter could be
$$ M(x) = \left( 1 + \frac {q}{1+x^2} \right)^2, \qquad q > 0, \eqno (18) $$ which has
yielded interesting results [18] with respect to the shape-invariant condition. In
(18), $q$ may be treated as a deformation parameter so that when $q \rightarrow 0$,
$M(x) \rightarrow 1$. The transformation (14) gives, because of (18), the relationship
$$ u(x) = x + q \tan^{-1} x. \eqno (19) $$ It implies that as $x \rightarrow \pm
\infty$, a small deformation has an insignificant effect on the variable $x$.

\noindent (b) When $F^2 \not= 1$, we can eliminate $M(x)$ to obtain from (6)
$$ F^2 + \delta G^2= 1, \eqno (20) $$ where $\delta$ is subject to the restriction
$\delta > 0$ if $F^2 < 1$ and $\delta < 0$ if $F^2 > 1$. The case $F^2 = 1$ is
incorporated for $\delta = 0$. Consistency with (17) demands that we set $\delta =
\frac {1}{b^2}$ for $F^2 < 1$ and $\delta = -\frac {1}{b^2}$ for $F^2 > 1$.

We can employ the machinery [24] of $so (2,1)$ to determine the wave functions $\chi
(\equiv \chi_{k \mu})$. First the operator relation $J_- \chi_0 e^{ik\phi} = 0$ is
solved for $\chi_0 =
\chi_{kk}$ and then $\chi_n = \chi_{k,k+n}$ are calculated for $n = 1,2,\ldots$ by
evaluating $J_+^n \chi_0 e^{ik\phi}$. To obtain solutions for the same potential
$V_k$ we need to replace $k$ by
$k-1, k-2, \ldots$ in the expressions for $\chi_1, \chi_2, \ldots$, respectively. We
then arrive at a chain of solutions
$$ \chi_0 \sim G^{k-\frac {1}{2}} \exp \left( \int \sqrt {M} G dx \right), $$
$$ \chi_1 \sim \left[ G - (k-1) F\right] G^{k-\frac {3}{2}}
\exp \left( \int \sqrt {M} G dx \right), \eqno (21)$$
$$ \chi_2 \sim \left\{ 2\left[ G - (k-1) F \right] \left[ G - (k-2) F \right] -
(k-2) \right\} G^{k - \frac {5}{2}} \exp \left( \int \sqrt {M} G dx \right), $$ and in a 
similar way the higher ones. These eigenfunctions correspond to the same
potential $V_k$, given by (9) for $\mu = k$. In view of (20), $V_k$ can be expressed as
$$ V_k = \left[ 1 + \delta \left(\frac {1}{4} - k^2\right) \right] G^2 - 2k FG. \eqno
(22)
$$ Corresponding to the three classes of solutions (17) we obtain specifically
$$ F^2 < 1 :\qquad V_k = \left(b^2 -k^2 + \frac {1}{4}\right) \sech^2 (u-c) - 2kb 
\sech (u-c) \tanh (u-c), $$
$$ F^2  = 0 :\qquad V_k = b^2 e^{\mp 2u} \mp 2kb e^{\mp u}, \eqno (23)$$
$$ F^2 > 1 :\qquad V_k = \left(b^2 + k^2 - \frac {1}{4}\right) \cosech^2 (u-c) - 2kb 
\cosech (u-c) \coth (u-c),  $$ where $u$ may depend on $x$ as in (19). The
three potentials above can be looked upon as mass-deformed versions of Scarf II, Morse
and generalized P\"oschl-Teller ones, respectively. The accompanying eigenfunctions are
obtained from (21) with the corresponding substitution of the functions $F$ and $G$.

{}Finally let us demonstrate that $V_{\rm eff}$ in (11) for $\mu= k$ coincides with the
intertwining-led effective potential provided by the condition $\eta H= H_1 \eta$,
where the intertwining operator
$\eta$ is taken in the first-order form $\eta = A(x) \frac {d}{dx} + B(x)$ [21]. The
intertwining relation implies that if $E_n$ $(n = 0,1,2,\ldots)$ are the bound-state
eigenvalues of $H$ then those of $H_1$, having an associated potential $V_{1, \rm
eff}$, are
$E_{1,n} = E_{n+1}$ if $\eta \psi_0 = 0$, where $\psi_0$ is the ground-state
eigenfunction of
$H$. Plugging in $H$ from (3) we get 
$$ A(x) = M^{-1/2}, \eqno (24) $$
$$ V_{\rm eff} (x) = \lambda + B^2 - (AB)^\prime. \eqno (25) $$ Note that
$V_{1,\rm eff} (x)$ depends upon $V_{\rm eff} (x)$ through the relation
$V_{1,\rm eff} (x) = V_{\rm eff} + 2AB^\prime - AA^{\prime\prime}$. In (25),
$\lambda$ denotes some integration constant.

The effective potentials given by the expression (11) for $\mu = k$ and by equation
(25) coincide for $B(x)$ given by
$$ B(x) = -\frac {M^\prime}{4M^{3/2}} + f(x), \eqno (26) $$ provided $f$ satisfies
$$ f^2 - \frac {1}{\sqrt{M}} f^\prime = V_k - \lambda. \eqno (27) $$

Equation (27) can be readily solved by using $V_k$ in (22) and looking for solutions
of the type $f = \zeta F + \sigma G$, where $\zeta$ and $\sigma$ are two constants to
be determined. We obtain the following class of solutions
$$ f_\pm (x) = \left(\pm k - \frac {1}{2}\right) F \mp G, \qquad \lambda_\pm = -
\left(k \mp \frac {1}{2}\right)^2, \eqno (28) $$ where $F$ and $G$ are given by any
one of the set (17). Thus the intertwining approach is consistent with the $so(2,1)$
algebra for the above solutions of $f$ and $\lambda$.

To conclude, we explored the properties of $so (2,1)$ in the context of PDEM
Schr\"odinger equation and generated mass-deformed versions of the Scarf II, Morse
and generalized P\"oschl-Teller potentials. We also sought consistency with the
intertwining condition and obtained the associated class of solutions.

\section*{Acknowledgements}

Two of us (BB and RR) gratefully acknowledge the support of the National Fund for
Scientific Research (FNRS), Belgium and thank Prof. C. Quesne for warm hospitality at
PNTPM, Universit\'e Libre de Bruxelles. One of us (BB) acknowledges useful
conversations with Prof. M. Berry and Prof. M. Znojil and thanks the International
Centre for Theoretical Physics, Trieste and \'Ustav jadern\'e fyziky, Czech Republic
for short term visits. PG is grateful to the Council of Scientific and Industrial
Research (CSIR), New Delhi for the award of a fellowship. CQ is a Research Director of
the National Fund for Scientific Research (FNRS), Belgium.

\section*{References}

\begin{itemize}
\item [{1}] M. R. Geller, W. Kohn, Phys. Rev. Lett. 70 (1993) 3103.
\item [{2}] Ll. Serra, E. Lipparini, Europhys. Lett. 40 (1997) 667.
\item [{3}] M. Barranco, M. Pi, S. M. Gatica, E. S. Hern\'{a}ndez, J. Navarro, Phys.
Rev. B 56 (1997) 8997.
\item [{4}] P. Ring, P. Schuck, The Nuclear Many Body Problem, Springer-Verlag, New
York, 1980.
\item [{5}] F. Arias de Saavedra, J. Boronat, A. Polls, A. Fabrocini, Phys. Rev. B 50
(1994) 4248.
\item [{6}] A. Puente, Ll. Serra, M. Casas, Z. Phys. D 31 (1994) 283.
\item [{7}] L. Dekar, L. Chetouani, T. F. Hammann, J. Math. Phys. 39 (1998) 2551.
\item [{8}] L. Dekar, L. Chetouani, T. F. Hammann, Phys. Rev. A 59 (1999) 107.
\item [{9}] A. R. Plastino, A. Puente, M. Casas, F. Garcias, A. Plastino, Rev. Mex.
Fis. 46 (2000) 78.
\item [{10}] R. Ko\c{c}. M. Koca, E. K\"orc\"uk, J. Phys. A 35 (2002) L527.
\item [{11}] B. G\"on\"ul, O. \"Ozer, B. G\"on\"ul, F. \"Uzg\"un, Mod. Phys. Lett. A
17 (2002) 2453.
\item [{12}] A. D. Alhaidari, Phys. Rev. A 66 (2002) 042116.
\item [{13}] A. de Souza Dutra, C. A. S. Almeida, Phys. Lett. A 275 (2000) 25.
\item [{14}] A. de Souza Dutra, M. Hott, C. A. S. Almeida, Europhys. Lett. 62 (2003)
8.
\item [{15}] J. Yu, S.-H. Dong, G.-H. Sun, Phys. Lett. A 322 (2004) 290.
\item [{16}] B. Roy, P. Roy, J. Phys. A 35 (2002) 3961.
\item [{17}] V. Milanovi\'c, Z. Ikoni\'c, J. Phys. A 32 (1999) 7001.
\item [{18}] A. R. Plastino, A. Rigo, M. Casas, F. Garcias, A. Plastino, Phys. Rev. A
60 (1998) 4318.
\item [{19}] B. G\"on\"ul, B. G\"on\"ul, D. Tutcu, O. \"Ozer, Mod. Phys. Lett. A 17
(2002) 2057.
\item[{20}] C. Quesne, V. M. Tkachuk, J. Phys. A 37 (2004) 4267.
\item [{21}] B. Bagchi, P. Gorain, C. Quesne, R. Roychoudhury, A general scheme
for the effective-mass Schr\"odinger equation and the generation of the associated
potentials, Preprint quant-ph/0405193.
\item [{22}] O. von Roos, Phys. Rev. B 27 (1983) 7547.
\item [{23}] J. Wu and Y. Alhassid, J. Math. Phys. 31 (1990) 557.
\item [{24}] M. J. Englefield and C. Quesne, J. Phys. A 24 (1991) 3557.
\end{itemize}

\end{document}